\documentclass[aps,showpacs,preprint]{revtex4}
\usepackage{graphicx}
\usepackage{bm}
\usepackage{amsfonts,amsmath,amssymb}
\usepackage[squaren]{SIunits}

\begin{document}

\title{Impact of the Fe Doping on Magnetism in Perovskite Cobaltites}
\author{Xigang Luo, Wendong Xing, Zhaofeng Li, Gang Wu, and Xianhui Chen$^\ast$}
\affiliation{Hefei National Laboratory for Physical Science at
Microscale and Department of Physics, University of Science and
Technology of China, Hefei, Anhui 230026, People's Republic of
China}

\begin{abstract}
We systematically studied the magnetic and transport properties for
the polycrystalline samples of Fe-doped perovskite cobaltites:
Pr$_{1-y}$Ca$_{y}$Co$_{1-x}$Fe$_x$O$_3$ ($y$=0.3, $x$=0-0.15;
$y$=0.45, $x$=0-0.3) and
Gd$_{0.55}$Sr$_{0.45}$Co$_{1-x}$Fe$_x$O$_{3}$ ($x$=0-0.3). Fe doping
leads to an enhancement of the ferromagnetism in the systems of
Pr$_{1-y}$Ca$_{y}$Co$_{1-x}$Fe$_x$O$_3$, while the ferromagnetism is
suppressed with further increasing Fe content and spin-glass
behavior is observed at high doping level of Fe. In contrast, the
ferromagnetism is suppressed in the system
Gd$_{0.55}$Sr$_{0.45}$Co$_{1-x}$Fe$_x$O$_{3}$ as long as Fe is
doped, and no spin-glass behavior is observed in the sample with Fe
doping up to 0.3.  The competition between ferromagnetic
interactions through Fe$^{3+}$-O-(LS)Co$^{4+}$ and antiferromagnetic
interactions through Fe$^{3+}$-O-Fe$^{3+}$ and
Fe$^{3+}$-O-(IS)Co$^{3+}$ is considered to be responsible for the
behavior observed above. The average radius of the ions on A sites
plays the key role in determining what type of interactions Fe
doping mainly introduces.
\end{abstract}


\maketitle

\section{Introduction}
Transition-metal oxides with perovskite-type crystal structure
attracted a great deal of interests in the past two decades due to
the fascinating characters of superconducting, magnetic and
transport properties. These properties range from high-temperature
superconductivity,\cite{Lee} colossal
magnetoresistance,\cite{Salamon}
ferroelectricity,\cite{Ramesh,LeeMK}
multiferroics,\cite{Hur,Kimura,Pimenov} to co-incident
metal-insulator, structural, and magnetic phase
transitions.\cite{Imada} The perovskite cobaltites ACoO$_3$ was
discovered in 1950s\cite{Koehler,Heikes}. But they still attracted
interests due to a couple of unique properties; namely, the large
magnetoresistance (MR),\cite{Briceno} enormous Hall
effect,\cite{Samoilov,Baily} the existence of the spin-state
transitions,\cite{Imada,Kasper,Moritomo,Frontera,Tsubouchi1} and
the unusual magnetic ground states of doped
cobaltites.\cite{Wu,Kunhs,Hoch,Ghoshray}

The distinct feature in perovskite cobaltites compared to other
transition-metal oxides such as manganites is the existence of
various spin states of Co ions. Recent experimental and theoretical
investigations on perovskite cobaltites indicate that the spin
states are low-spin (LS) and the mixture of intermediate-spin(IS)/LS
for tetravalent and trivalent cobalt ions,
respectively.\cite{Lashkareva,Korotin,Yamaguchi1,Louca,Kobayashi,Zobel,Ravindran}
This behavior arises from the fact that the crystal field splitting
of Co $d$ states ($\Delta_{\rm CF}$) and the Hund's rule exchange
energy ($J_{\rm H}$) are comparable in magnitude for the cobaltites,
which means that the energy gap ($\delta E$ = $\Delta_{\rm
CF}$-$J_{\rm H}$) between $t_{\rm 2g}$ and $e_{\rm g}$ bands is
rather small. In fact, this gap can be of the order of 10 meV in
LaCoO$_3$, so that the electrons in $t_{\rm 2g}$ levels can be
thermally excited to the $e_{\rm g}$ states, leading to higher spin
states of Co ions.\cite{Imada} Because $\Delta_{\rm CF}$ is very
sensitive to the variation in the Co-O bond length ($d_{\rm Co-O}$)
or the unit cell volume of lattice, the subtle balance between
$\Delta_{\rm CF}$ and $J_{\rm ex}$ may be easily disrupted by
different kinds of effect, such as the hole-doping and the
chemical/external pressure.\cite{Asai,Lengsdorf,Vogt,Fita} The
$"$intermediate spin state$"$  has been claimed to exist in
La$_{1-x}$Sr$_x$CoO$_3$ by different
groups.\cite{Faith,Wu,Korotin,Ravindran} It is generally accepted
that the ferromagnetism in hole-doped La$_{1-x}$M$_{x}$CoO$_3$ (M =
Ca, Sr, and Ba) results from the double-exchange (DE) interaction
between Co$^{3+}$ and Co$^{4+}$ ions, facilitating also the
electrical conductivity in the ferromagnetic metallic phase. In this
consideration, the DE interaction is accomplished through transfer
of $e_{\rm g}$ electrons in the intermediate-spin (IS) Co$^{3+}$ to
LS Co$^{4+}$, so that the $e_{\rm g}$ electrons become collective in
La$_{1-x}$Sr$_x$CoO$_3$ and the Co ions ultimately turn into $t_{\rm
2g}^{5}e_{\rm g}^{x}$ electronic configuration. For the cobaltites
(such as: Pr$_{1-y}$Ca$_{y}$CoO$_3$ and Nd$_{1-y}$Ca$_{y}$CoO$_3$)
with small average radius of ions on A sites, resistivity is not
metallic in these compounds, meaning that the magnetic exchange
should be localized to some extent instead of being itinerant as in
DE exchange. Therefore, further studies to make clear the magnetic
exchange interactions in cobaltites are desired.

If Co ions are partially substituted by other transition metal ions,
the DE interaction between Co$^{3+}$ and Co$^{4+}$ will be destroyed
to some extent. In addition, magnetic exchange interactions between
Co ions and the doped transition-metal ions could be expected
because the existence of various valence and spin-states of Co ions.
Investigation on such interactions through substituting Co ions by
other transition ions would be expected helpfully to make the role
of all kinds of spin-states and valences in the magnetic exchange
interactions clear. In the present paper, we choose Fe element as
the substitution ion. The effect of Fe doping in
La$_{1-x}$Sr$_x$CoO$_3$ has been studied by several
groups,\cite{Sun,Czirak,Nemeth,Barman,Phan} and the common
conclusion is the suppression of ferromagnetism and metallicity with
Fe doping as a result of the diluting of ferromagnetic interaction
of Co$^{3+}$-O-Co$^{4+}$ and the introduction of the
antiferromagnetic exchange interactions Fe-O-Fe and Fe-O-Co.
However, Our recent work on
Pr$_{0.5}$Ca$_{0.5}$Co$_{1-x}$Fe$_x$O$_{3-\delta}$ gave the evidence
that Fe doping enhances the ferromagnetism.\cite{Luo1} The
contrasting results observed in La$_{1-x}$Sr$_x$CoO$_3$ and
Pr$_{0.5}$Ca$_{0.5}$CoO$_{3-\delta}$ with Fe doping could give some
constructive hints on the magnetic interactions and spin states in
the two systems. Because the existence of spin-state transition with
temperature in
Pr$_{0.5}$Ca$_{0.5}$CoO$_{3-\delta}$,\cite{Tsubouchi1,Luo1} it is
not suitable to be used for investigating the magnetic interactions
between Co and Fe ions. Therefore, we choose two systems with less
Ca content, which do not exhibit spin-state transition at low
temperature.\cite{Tsubouchi2} As a comparison, the system
Gd$_{0.55}$Sr$_{0.45}$CoO$_{3}$, which has the similar magnetic and
transport properties to the typical
La$_{1-x}$Sr$_x$CoO$_3$,\cite{Luo2} was chosen as another object for
Fe doping. The distinct responses of magnetic properties to the Fe
doping are observed due to the different average ionic radius of A
site.

\section{Experimental Details}
Polycrystalline Pr$_{1-y}$Ca$_{y}$Co$_{1-x}$Fe$_x$O$_3$ ($y$=0.3,
$x$=0-0.15; $y$=0.45, $x$=0-0.3) and
Gd$_{0.55}$Sr$_{0.45}$Co$_{1-x}$Fe$_x$O$_{3}$ ($x$=0-0.3) samples
were prepared through the conventional solid-state reaction. For
Pr$_{0.7}$Ca$_{0.3}$Co$_{1-x}$Fe$_x$O$_3$ and
Pr$_{0.55}$Ca$_{0.45}$Co$_{1-x}$Fe$_x$O$_3$ samples, the powders
of Pr$_6$O$_{11}$, CaCO$_3$, Co$_3$O$_4$, and Fe$_2$O$_3$ were
mixed in stoichiometric proportion and heated at 1200 $\celsius$
in the flowing oxygen for 24 h. The mixture was ground and pressed
into pellets and finally sintered  at 1200 $\celsius$ in flowing
oxygen for 24 h twice.
Gd$_{0.55}$Sr$_{0.45}$Co$_{1-x}$Fe$_x$O$_{3}$ samples were
fabricated from the stoichiometric amount of Gd$_2$O$_3$,
SrCO$_3$, and Co$_3$O$_4$ and Fe$_2$O$_3$ powders. The mixtures
were fired at 1200 $\celsius$ for 24 h in air, then reground and
pressed into pellets which were sequently sintered at 1200
$\celsius$ for 24 h. This procedure was then repeated two times.
Because in these as-fabricated cobaltites a large amount of oxygen
vacancies exist,\cite{Luo1,Luo2} the samples need to be
post-annealed under high pressure oxygen to achieve oxygen
stoichiometry. Pr$_{1-y}$Ca$_{y}$Co$_{1-x}$Fe$_x$O$_3$ samples
were then annealed at 600 $\celsius$ under the high oxygen
pressure of 265 atm for 48 h. The
Gd$_{0.55}$Sr$_{0.45}$Co$_{1-x}$Fe$_x$O$_{3}$ samples were
annealed at 900 $\celsius$ under the high oxygen pressure of 260
atm for 24 h.

X-ray diffraction (XRD) data were collected using Cu $K_\alpha$
radiation ($\lambda$ = 1.5418${\rm \AA}$) at room temperature.
Direct and alternating current (dc and ac) magnetic measurements
were performed with a superconducting quantum interference device
(SQUID) magnetometer (MPMS-7XL, Quantum Design). Resistivity was
measured by using the standard ac four-probe method. The oxygen
content of the samples were determined by using the K$_2$Cr$_2$O$_7$
titration method.\cite{Luo1,Luo2} The results indicated that after
annealing under the high pressure oxygen the oxygen vacancies are
controlled less than 0.01 for the
Gd$_{0.55}$Sr$_{0.45}$Co$_{1-x}$Fe$_x$O$_{3}$ samples and 0.005 for
the Pr$_{1-y}$Ca$_{y}$Co$_{1-x}$Fe$_x$O$_3$ samples.

\section{Experimental Results}
\subsection{Structural Characterization}
XRD patterns indicated that all the samples are single phase with
the orthorhombic structural symmetry (Space Group: Pnma),
consistent with our previous reports.\cite{Luo1,Luo2} The unit
cell volumes with Fe concentration are plotted in Fig.1 for the
three systems. It indicated that for all the three systems,
Fe-doping enlarges the crystal lattice obviously. As one knows,
the radius of LS Co$^{4+}$, LS Co$^{3+}$, IS Co$^{3+}$, and
high-spin (HS) Co$^{3+}$ is 0.53, 0.545, 0.56 and 0.61 ${\rm
\AA}$, respectively; while the radius of LS Fe$^{3+}$, HS
Fe$^{3+}$ and Fe$^{4+}$ is 0.55, 0.645 and 0.585 ${\rm \AA}$,
respectively.\cite{Shannon} Consequently, the rapid enlargement of
unit cell volume shown in Fig. 1 strongly suggested that the doped
Fe ions is Fe$^{3+}$ ions with a high-spin (HS) $t_{\rm
2g}^{3}e_{\rm g}^{2}$ electronic configuration, consistent with
the results obtained by the M${\rm \ddot{o}}$ssbauer experiments
in the other Fe-doped perovskite cobaltites TbBaCo$_2$O$_{5.5}$
and La$_{1-x}$Sr$_{x}$CoO$_{3}$.\cite{Kopcewicz,Czirak,Homanay}

\subsection{Magnetic properties}

Temperature dependence of zero-field cooled (ZFC) and field-cooled
(FC) molar magnetization ($H$ = 0.1 T) is shown in Fig. 2 for
Gd$_{0.55}$Sr$_{0.45}$Co$_{1-x}$Fe$_x$O$_{3}$ samples. Fig. 2 shows
that the ferromagnetism is suppressed as any Co ions are substituted
by Fe ions. The Fe-free sample has $T_{\rm c}$ $\approx$ 125 K
($T_{\rm c}$ is determined from the maximum of the FC d$M(T)$/d$T$),
while for the sample with $x$ = 0.30, $T_{\rm c}$ decreases to about
95 K. The magnetization at 4 K is also reduced from 10500 emu/mol
for $x$ = 0 to 3000 emu/mol for $x$ = 0.3 except for the enhancement
for $x$ = 0.1. The effect of Fe doping in
Gd$_{0.55}$Sr$_{0.45}$CoO$_{3}$ seems to be the same as that in
La$_{1-x}$Sr$_x$CoO$_3$.\cite{Sun,Phan} In La$_{1-x}$Sr$_x$CoO$_3$,
ferromagnetism are suppressed when Co ions are substituted by Fe
ions, and spin-glass behavior is induced with further increasing Fe
concentration.\cite{Phan} However, no spin-glass behavior can be
observed in Gd$_{0.55}$Sr$_{0.45}$Co$_{1-x}$Fe$_{x}$O$_{3}$ even
with x up to 0.3. It is confirmed from the ac susceptibility
measurements. Temperature dependence of in-phase ac susceptibility
($\chi~'(T)$) is shown in Fig. 3 for three samples with $x$ = 0,
0.15 and 0.3. $\chi'(T)$ is measured in an ac magnetic field of $H$
= 3.8 Oe at 1 and 1000 Hz, respectively. Strong peaks can be
observed for all the $\chi~'(T)$ curves around the temperature where
the d$M$/d$T$ reaches maximum, suggesting that the peaks in the
$\chi~'(T)$ curves correspond to the ferromagnetic transition. The
position of the peak  at different frequencies shifts less than 1 K
($\leq~1\%$). Therefore, the frequency independent-$\chi~'(T)$
suggests  no spin-glass behavior even for the sample with $x$=0.30,
consistent with that observed in the $M(T)$ curves. This behavior is
in contrast to the case of La$_{1-x}$Sr$_x$Co$_{1-x}$Fe$_{x}$O$_3$,
in which a spin glass state is induced in the sample with large
x.\cite{Phan} The very weak shift of the peak position with
frequency could arise from the cluster nature of the ferromagnetism
in the perovskite cobaltites.\cite{Ghoshray}

More intriguing magnetism with Fe doping is observed in
Pr$_{1-y}$Ca$_{y}$CoO$_{3}$ ($y$ = 0.3 and 0.45) systems. Figure 4
shows temperature dependence of the ZFC and FC molar magnetization
for the Pr$_{1-y}$Ca$_{y}$Co$_{1-x}$Fe$_x$O$_3$ ($y$=0.3,
$x$=0-0.15; $y$=0.45, $x$=0-0.3) samples. Ferromagnetic transition
occurs at $T_{\rm c}$ $\approx$ 69 K for $y$ = 0.45 and 46 K for $y$
= 0.3 in the Fe-free samples, respectively. With Fe doping,
 $T_{\rm c}$ increases firstly and then decreases for the two series of samples with different $y$.
  It suggests that
ferromagnetism in the two series of samples is enhanced firstly, and
then suppressed with further increasing Fe concentration. For the
highly doped samples
Pr$_{0.55}$Ca$_{0.45}$Co$_{0.7}$Fe$_{0.3}$O$_{3}$ and
Pr$_{0.7}$Ca$_{0.3}$Co$_{0.85}$Fe$_{0.15}$O$_{3}$,  spin-glass
behavior is observed, being in contrast to the case of
Gd$_{0.55}$Sr$_{0.45}$Co$_{1-x}$Fe$_{x}$O$_{3}$ samples.

To clarify the magnetic nature of the data in Fig.4, the ZFC
$\chi~'(T)$ was measured in the magnetic field of $H$ = 3.8 Oe at 1
and 1000 Hz for all the samples. The results are shown in Fig.5 and
Fig.6, respectively. It is found that the position of the peaks is
only 2-3 K different from the temperature of maximum of d$M$/d$T$
($T_{\rm c}$) for the slightly doped samples, while more than 10 K
for the highly doped samples (15 K for
Pr$_{0.55}$Ca$_{0.45}$Co$_{0.7}$Fe$_{0.3}$O$_{3}$ and 11 K for
Pr$_{0.7}$Ca$_{0.3}$Co$_{0.85}$Fe$_{0.15}$O$_{3}$). Frequency
dependence of the peak position is almost independent for the
slightly doped samples (less than 1 K in Fig.5(a-d) and Fig.6(a-c),
while  the peak position is strongly frequency-dependent for the
samples with $x = 0.3$ for $y = 0.45$ and $x\geq0.1$ for $y = 0.3$,
and  the temperature of peaks shifts larger than 5$\%$ at two
different frequencies of 1 Hz and 1 KHz as shown in Fig.5e and
Fig.6e, indicative of the obvious spin-glass behavior. The frequency
dependence is a direct indication of the slow spin dynamics,
indicating the peaks in Fig.5e and Fig.6e associated with the
spin-glass freezing temperature $T_f$. The frequency dependence is
illustrated more clearly in Fig.7, it shows the $"$closeup$"$ of the
peaks in $\chi~'(T)$ shown in Fig.5e. It is found that the $T_f$
increases monotonically with increasing frequency $f$, further
confirming the spin-glass state in
Pr$_{0.55}$Ca$_{0.45}$Co$_{0.7}$Fe$_{0.3}$O$_{3}$. The frequency
dependence can be well described by the conventional critical
$"$slowing down$"$ of the spin dynamics, \cite{Gunnarson,Mydosh} as
described by
\begin{eqnarray}
\frac{\tau}{\tau_0} \propto {(\frac{T_f-T_{SG}}{T_{SG}})}^{-z\nu}
\end{eqnarray}
where $\tau_{0}$ is the characteristic time scale for the spin
dynamics(i.e., the relaxation time), $\tau$=$f^{-1}$, $T_{SG}$ is
the critical temperature for spin-glass ordering (this is equivalent
to the  value of $T_f$ when $f\longrightarrow$0) and $z\nu$ is a
constant exponent (in which $\nu$ is the critical exponent of the
correlation length and $z$ the dynamic exponent). The best fit to
the data using the Eq. (1) is obtained by choosing the value of
$T_{SG}$ which minimizes the least-square deviation from a
straight-line fit (See Fig.8). The values of $\tau_{0}$ and $z\nu $
are extracted from the intercept and slope, respectively. For
Pr$_{0.55}$Ca$_{0.45}$Co$_{0.7}$Fe$_{0.3}$O$_{3}$, $T_{SG}$ = 51.81
K, $z\nu $ = 6.85, and $\tau_{0}$ = 1.37$\times$10$^{-12}$ s, which
is larger than that of conventional spin glasses ($\sim$ 10$^{-13}$
s), indicating a slower spin dynamics. $\tau_{0}$ is larger than
that obtained in La$_{1-x}$Sr$_{x}$CoO$_{3}$ ($x$ $<$
0.18),\cite{Wu} but smaller than that achieved in Fe-doped
La$_{0.5}$Sr$_{0.5}$CoO$_{3}$.\cite{Phan}

\subsection{Transport Behaviors}
Although the effect of Fe-doping on the magnetism is different for
Gd$_{0.55}$Sr$_{0.45}$CoO$_{3}$ and Pr$_{1-y}$Ca$_{y}$CoO$_{3}$
($y$ = 0.3 and 0.45) systems, the evolution of resistivity with Fe
doping shows similar. Temperature dependence of the resistivity
for the three series of samples is shown in Fig. 9. The zero field
data indicate that the Gd$_{0.55}$Sr$_{0.45}$CoO$_{3}$ is metallic
(d$\rho$/d$T >$ 0) in the whole measuring temperature range, while
semiconducting behavior (d$\rho$/d$T <$ 0) is observed above
$T_{\rm c}$ for Pr$_{1-y}$Ca$_{y}$CoO$_{3}$ ($y$ = 0.3 and 0.45)
samples. The evolution of resistivity with Fe doping is similar
among all the three systems: it is found that the resistivity
increases monotonically with increasing Fe doping level; the
highly Fe-doped samples show an insulating behavior in the whole
temperature range. These results are similar to those observed in
La$_{0.5}$Sr$_{0.5}$Co$_{1-x}$Fe$_{x}$O$_{3}$ samples.\cite{Phan}
Resistivity results suggest that Fe doping leads to strong
localized feature, and the conductive channels between Co ions are
broken near Fe atoms.

\section{Discussions}
The transport and magnetization data indicate that the
Gd$_{0.55}$Sr$_{0.45}$CoO$_{3}$ has much higher ferromagnetic
transition temperature relative to Pr$_{1-y}$Ca$_{y}$CoO$_{3}$.
However, Gd$_{0.55}$Sr$_{0.45}$CoO$_{3}$ has much lower
ferromagnetic transition temperature relative to the same Sr doping
level in La$_{1-x}$Sr$_{x}$CoO$_3$.\cite{Luo1} The ferromagnetic
transition strongly depends on the average radius of ions on A sites
($<r_{\rm A}>$), so that the La-, Pr- and Gd-based systems show
different ferromagnetic transition temperature ( $<r_{\rm
La0.55Sr0.45}$$>$ = 1.258 ${\rm \AA}$, $<r_{\rm Gd0.55Sr0.45}$$>$ =
1.198 ${\rm \AA}$, $<r_{\rm Pr0.55Ca0.45}$$>$ = 1.179 ${\rm
\AA}$)\cite{explain}. This is because the average radius of ions on
A sites ($<r_{\rm A}>$) strongly affects the band width
\cite{Yamaguchi2}, consequently on $T_{\rm c}$. In general, band
width and $T_{\rm c}$ decreases with decreasing $<r_{\rm A}>$. The
band width directly influences the transport behavior. As shown in
Fig.9, the Fe-free Gd$_{0.55}$Sr$_{0.45}$CoO$_{3}$ sample shows
metallic behavior in the whole temperature range, similar to that of
La$_{1-x}$Sr$_{x}$CoO$_3$. It suggests that holes should be
itinerant, so that $e_g$ is considered to be collective.
Consequently, most of the Co ions (except for some LS trivalent Co
ions) in Gd$_{0.55}$Sr$_{0.45}$CoO$_{3}$ have intermediate $t_{\rm
2g}^5e_{\rm g}^\delta$ electronic configuration. However,
Pr$_{0.55}$Ca$_{0.45}$CoO$_{3}$ shows weakly metallic behavior only
above 250 K and becomes insulating below 250 K. The different
temperature dependence of resistivity in
Gd$_{0.55}$Sr$_{0.45}$CoO$_{3}$ and Pr$_{0.55}$Ca$_{0.45}$CoO$_{3}$
could be understood with the different $<r_{\rm A}>$. Larger
$<r_{\rm A}>$ in Gd$_{0.55}$Sr$_{0.45}$CoO$_{3}$ leads to a larger
band width, so that it is metallic below room temperature; while
smaller $<r_{\rm A}>$ in Pr$_{0.55}$Ca$_{0.45}$CoO$_{3}$ results in
a narrower band width, so that it shows insulating behavior in
$T_{\rm c} < T <$ 250K, indicative of the obvious localized feature.

M${\rm \ddot{o}}$ssbauer
experiments\cite{Kopcewicz,Czirak,Homanay} have demonstrated that
the doped Fe ions in the perovskite cobaltites have the formation
of Fe$^{3+}$ with a high-spin $t_{\rm 2g}^3e_{\rm g}^2$ electronic
configuration. The rapid enlargement of lattice volume with Fe
doping shown in Fig.1 further confirms that the Fe ions have such
formation in the present three series of samples. Recent
investigations support the configuration of LS Co$^{4+}$ and the
mixed IS/LS Co$^{3+}$ in perovskite
cobaltites.\cite{Lashkareva,Korotin,Yamaguchi1,Louca,Kobayashi,Zobel,Ravindran}
By fitting the susceptibility above $T_{\rm c}$ with the
Curie-Weiss law after subtracting the contribution of Gd$^{3+}$
(7.94$\mu_B$)/Pr$^{3+}$(3.58$\mu_B$) ions, the effective moment
per Co ion is obtained as 2.362$\mu_B$ and 2.368$\mu_B$  for
Gd$_{0.55}$Ca$_{0.45}$CoO$_{3}$ and
Pr$_{0.55}$Ca$_{0.45}$CoO$_{3}$, respectively. These values are
close to that (2.398$\mu_B$) estimated from IS Co$^{3+}$ + LS
Co$^{4+}$ configuration for such doping level, indicating that
Co$^{4+}$ ions are at LS state and most of the Co$^{3+}$ ions are
at IS state in these two compounds. Therefore, the exchange
interactions including Fe$^{3+}$ are mainly Fe$^{3+}$-O-Fe$^{3+}$,
Fe$^{3+}$-O-Co$^{3+}$(IS), Fe$^{3+}$-O-Co$^{4+}$(LS) as shown in
Fig.10. According to Goodenough-Kanamori rules, superexchange
interactions through Fe$^{3+}$-O-Fe$^{3+}$ are antiferromagnetic,
while those through Fe$^{3+}$-O-Co$^{4+}$(LS) are
ferromagnetic.\cite{Goodenough1955,Kanamori1959,Goodenough1963}
The Goodenough-Kanamori rules can not directly give whether the
exchange interactions through Fe$^{3+}$-O-Co$^{3+}$(IS) are
ferromagnetic or antiferromagnetic because the sign of this
interaction depends on the relative orientation of the
(un-)occupied $e_{\rm g}$ orbitals. Although the information of
the relative orientation of the orbitals in present materials is
not available, we could assume an antiferromagnetic exchange
interaction through Fe$^{3+}$-O-Co$^{3+}$(IS) based on the
observed results in
Gd$_{0.55}$Sr$_{0.45}$Co$_{1-x}$Fe$_{x}$O$_{3}$. In
Gd$_{0.55}$Sr$_{0.45}$CoO$_{3}$, the exchange interactions through
Fe$^{3+}$-O-Co$^{4+}$(LS) are absent due to the collective feature
of the $e_{\rm g}$ electrons (forming the intermediate $t_{\rm
2g}^5e_{\rm g}^\delta$ electronic configuration instead of
individual LS Co$^{4+}$ and IS Co$^{3+}$ ions). At the low doping
level of Fe, Fe ions is diluted so that the interactions through
Fe$^{3+}$-O-Co$^{3+}$(IS) play the main role. Therefore the
suppression of the ferromagnetism by slightly doping of Fe
suggests that the interactions through Fe$^{3+}$-O-Co$^{3+}$(IS)
should be antiferromagnetic. With increasing Fe concentration, the
strong antiferromagnetic exchange interactions through
Fe$^{3+}$-O-Fe$^{3+}$ appear and further suppress the
ferromagnetism . In Pr$_{1-y}$Ca$_{y}$CoO$_{3}$, the $e_{\rm g}$
electrons have obviously the localized feature as discussed above,
so that the Co ions are at LS Co$^{4+}$ and IS Co$^{3+}$ states.
For the slightly doping samples, the enhancement of ferromagnetism
arises from the ferromagnetic interactions through
Fe$^{3+}$-O-Co$^{4+}$(LS). The enhancement of the ferromagnetism
at low Fe doping level suggests that ferromagnetic
Fe$^{3+}$-O-Co$^{4+}$(LS) interaction is stronger than possibly
antiferromagnetic Fe$^{3+}$-O-Co$^{3+}$(IS) one, so that the
ferromagnetic Fe$^{3+}$-O-Co$^{4+}$(LS) interactions are expected
to play the main role in the slight Fe-doped samples. However,
with further increasing Fe concentration, the interactions through
Fe$^{3+}$-O-Fe$^{3+}$ increase dramatically. According to
Goodenough-Kanamori rules, the exchange interactions through
Fe$^{3+}$-O-Fe$^{3+}$ is stronger than those through
Fe$^{3+}$-O-Co$^{4+}$(LS).\cite{Goodenough1963} In this case, the
antiferromagnetic Fe$^{3+}$-O-Fe$^{3+}$ interactions gradually
become dominating.

It is well known that the competition between ferromagnetic and
antiferromagnetic interactions as well as the randomness of magnetic
interactions induces the spin-glass magnetism. The antiferromagnetic
interactions (through Fe$^{3+}$-O-Co$^{3}$(IS) and
Fe$^{3+}$-O-Fe$^{3+}$) introduced by Fe doping coexist with the
ferromagnetic interactions between Co ions (as well as
Fe$^{3+}$-O-Co$^{4+}$(LS)). At the meantime, the disordered
distribution of Fe ions on Co sites leads to the strong randomness
of (anti)ferromagnetic interactions in the system, consequently
induces the magnetic frustration. The two effects induced by Fe
doping induce the spin-glass magnetism with increasing Fe
concentration. In Pr$_{0.55}$Ca$_{0.45}$Co$_{1-x}$Fe$_{x}$O$_{3}$
and Pr$_{0.7}$Ca$_{0.3}$Co$_{1-x}$Fe$_{x}$O$_{3}$, the spin-glass
behavior is observed in the highly doped samples. It should be
pointed out that spin-glass behavior occurs at lower Fe doping level
for Pr$_{0.7}$Ca$_{0.3}$Co$_{1-x}$Fe$_{x}$O$_{3}$ than for
Pr$_{0.55}$Ca$_{0.45}$Co$_{1-x}$Fe$_{x}$O$_{3}$. This should arise
from the weaker ferromagnetism in Pr$_{0.7}$Ca$_{0.3}$CoO$_{3}$. It
is strange that spin-glass behavior is not found in
Gd$_{0.55}$Sr$_{0.45}$Co$_{0.7}$Fe$_{0.3}$O$_{3}$ (see Fig.2 and
Fig.3c) although $T_{\rm c}$ decreases by 30 K with Fe doping of
0.30 in Gd$_{0.55}$Sr$_{0.45}$CoO$_{3}$. This is in contrast to that
observed in Fe-doped La$_{1-x}$Sr$_{x}$CoO$_3$ system, in which the
strong spin-glass behavior is also found with the Fe doping level up
to 0.30.\cite{Sun,Phan} Previous research\cite{Luo2} in
Gd$_{1-y}$Sr$_{y}$CoO$_3$ have demonstrated that there is not
spin-glass behavior at low Sr concentration, being in contrast to
the spin-glass magnetism in La$_{1-x}$Sr$_{x}$CoO$_3$ ($x< 0.18$).
The existence of Gd$^{3+}$ ions with the {\sl large magnetic moment}
(S = 7/2) may be taken as one possible reason for the absence of
spin-glass behavior in Gd$_{1-y}$Sr$_{y}$CoO$_3$ because the strong
internal field produced by Gd$^{3+}$ magnetic sublattice may reduce
the relaxation time and destroy a spin-glass magnetism.

\section{Conclusion}
We systematically studied the effect of Fe doping on the magnetic
and transport properties on the polycrystalline
Pr$_{0.7}$Ca$_{0.3}$Co$_{1-x}$Fe$_x$O$_{3-\delta}$,
Pr$_{0.55}$Ca$_{0.45}$Co$_{1-x}$Fe$_x$O$_{3-\delta}$ and
Gd$_{0.55}$Sr$_{0.45}$Co$_{1-x}$Fe$_x$O$_{3}$. It is found that that
Fe doping in Gd$_{0.55}$Sr$_{0.45}$CoO$_{3}$ strongly suppresses
ferromagnetism, but does not induce spin-glass behavior; while Fe
doping in Pr$_{1-y}$Ca$_{y}$CoO$_{3}$ enhances the ferromagnetism at
low Fe doping level, but suppresses ferromagnetism and induced
spin-glass magnetism in highly doped samples. Such contrasting
response to Fe doping in Gd$_{0.55}$Sr$_{0.45}$CoO$_{3}$ and
Pr$_{1-y}$Ca$_{y}$CoO$_{3}$ can be interpreted on the basis of the
existences of the antiferromagnetic interactions through
Fe$^{3+}$-O-Fe$^{3+}$ and Fe$^{3+}$-O-Co$^{3+}$(IS) and
ferromagnetic interactions through Fe$^{3+}$-O-Co$^{4+}$(LS). In
this picture, the average radius of the ions on A sites plays the
key role in determining what type interaction Fe doping mainly
introduces.

\section{Acknowledgement}
This work is supported by the Nature Science Foundation of China and
by the Ministry of Science and Technology of China (973 project No:
2006CB601001), and by the Knowledge Innovation Project of Chinese
Academy of Sciences.\\[3mm]

 $\ast$ Electronic address: chenxh@ustc.edu.cn

\clearpage

\begin{figure}
\centering
\includegraphics[width=0.75\textwidth]{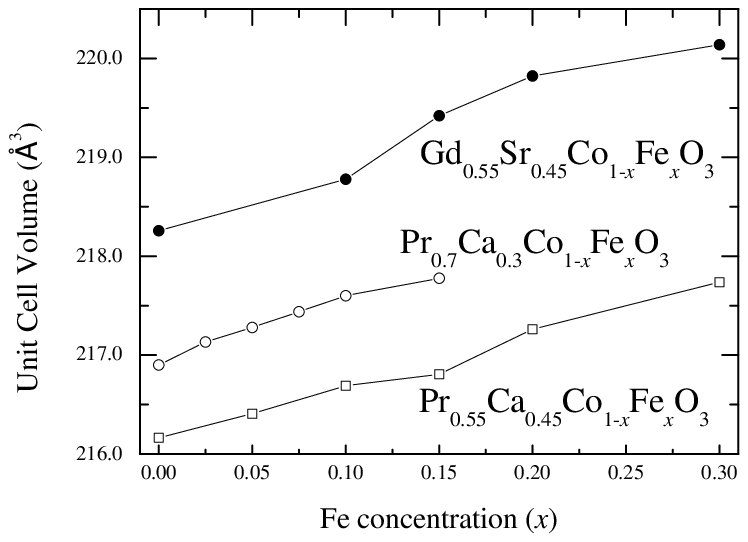}
\caption{Variation of unit cell volume with $x$ for the
polycrystalline Pr$_{1-y}$Ca$_{y}$Co$_{1-x}$Fe$_x$O$_3$, and
Gd$_{0.55}$Sr$_{0.45}$Co$_{1-x}$Fe$_{x}$O$_{3}$.}
\end{figure}

\begin{figure}
\centering
\includegraphics[width=0.75\textwidth]{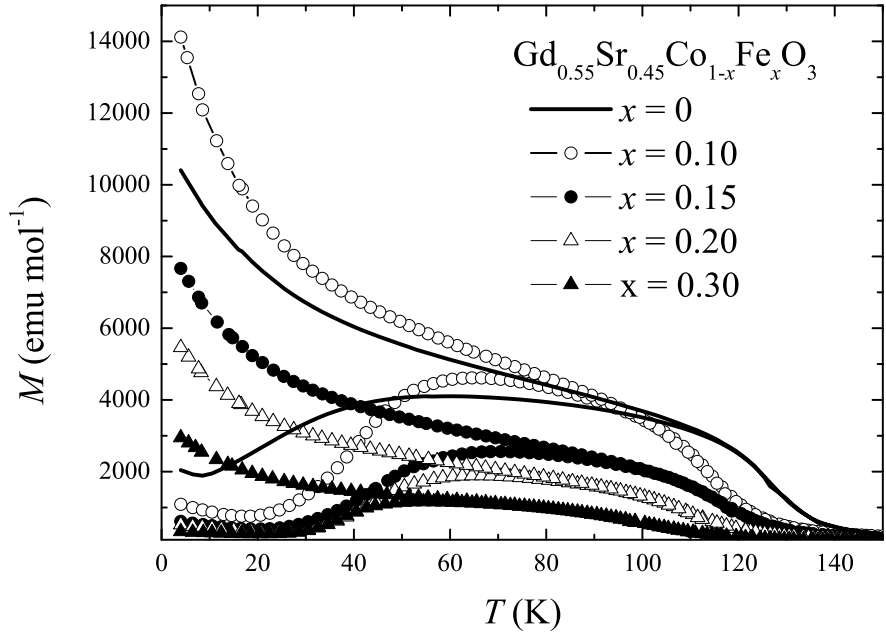}
\caption{Temperature dependence of the molar magnetization recorded
at $H$ = 0.1 T for the polycrystalline samples
Gd$_{0.55}$Sr$_{0.45}$Co$_{1-x}$Fe$_{x}$O$_{3}$. }
\end{figure}

\begin{figure} \centering
\includegraphics[width=0.6\textwidth]{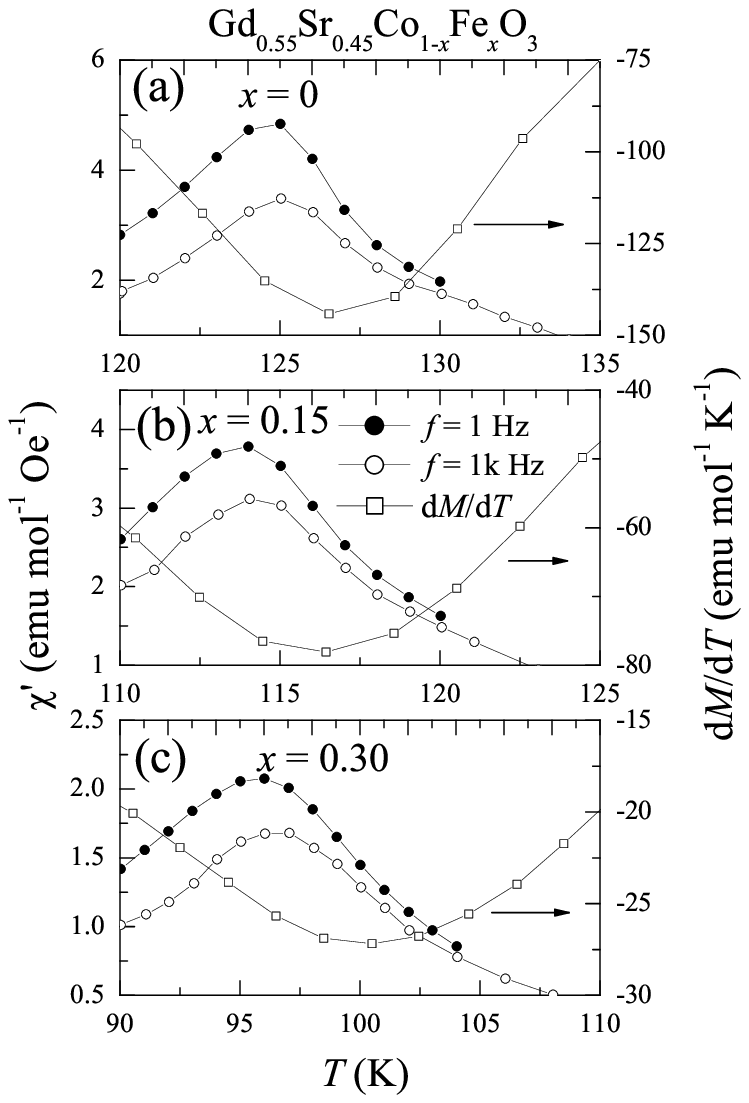}
\caption{Temperature dependence of in-phase ac susceptibility
$\chi'(T)$ at 1 Hz and 1 kHz measured in ac the magnetic field of
3.8 Oe for Gd$_{0.55}$Sr$_{0.45}$Co$_{1-x}$Fe$_{x}$O$_{3}$ with $x$
= 0, 0.1,and 0.3. As a comparison to dc data, the d$M$/d$T$ is
plotted together with the ac data.}
\end{figure}

\begin{figure}
\centering
\includegraphics[width=0.9\textwidth]{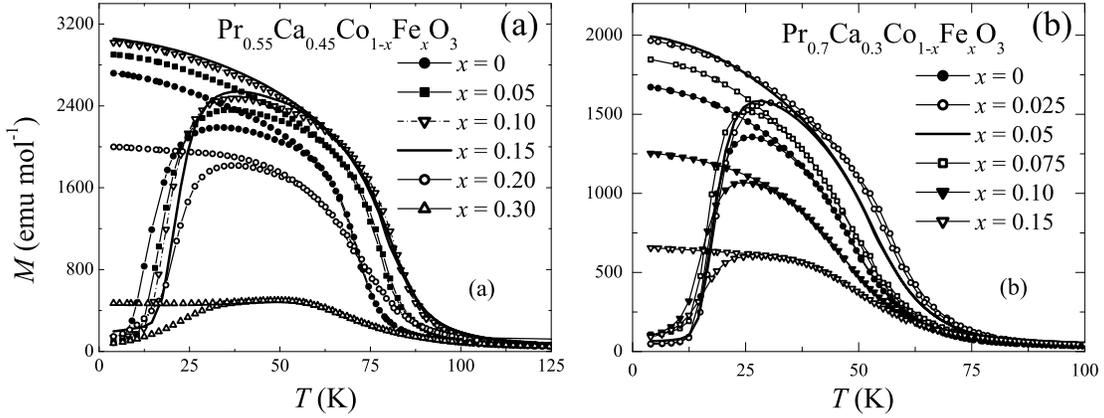}
\caption{Temperature dependence of the molar magnetization recorded
at $H$ = 0.1 T for the polycrystalline samples: (a)
Pr$_{0.55}$Ca$_{0.45}$Co$_{1-x}$Fe$_{x}$O$_{3}$ and (b)
Pr$_{0.7}$Ca$_{0.3}$Co$_{1-x}$Fe$_{x}$O$_{3}$.}
\end{figure}

\begin{figure} \centering
\includegraphics[width=0.75\textwidth]{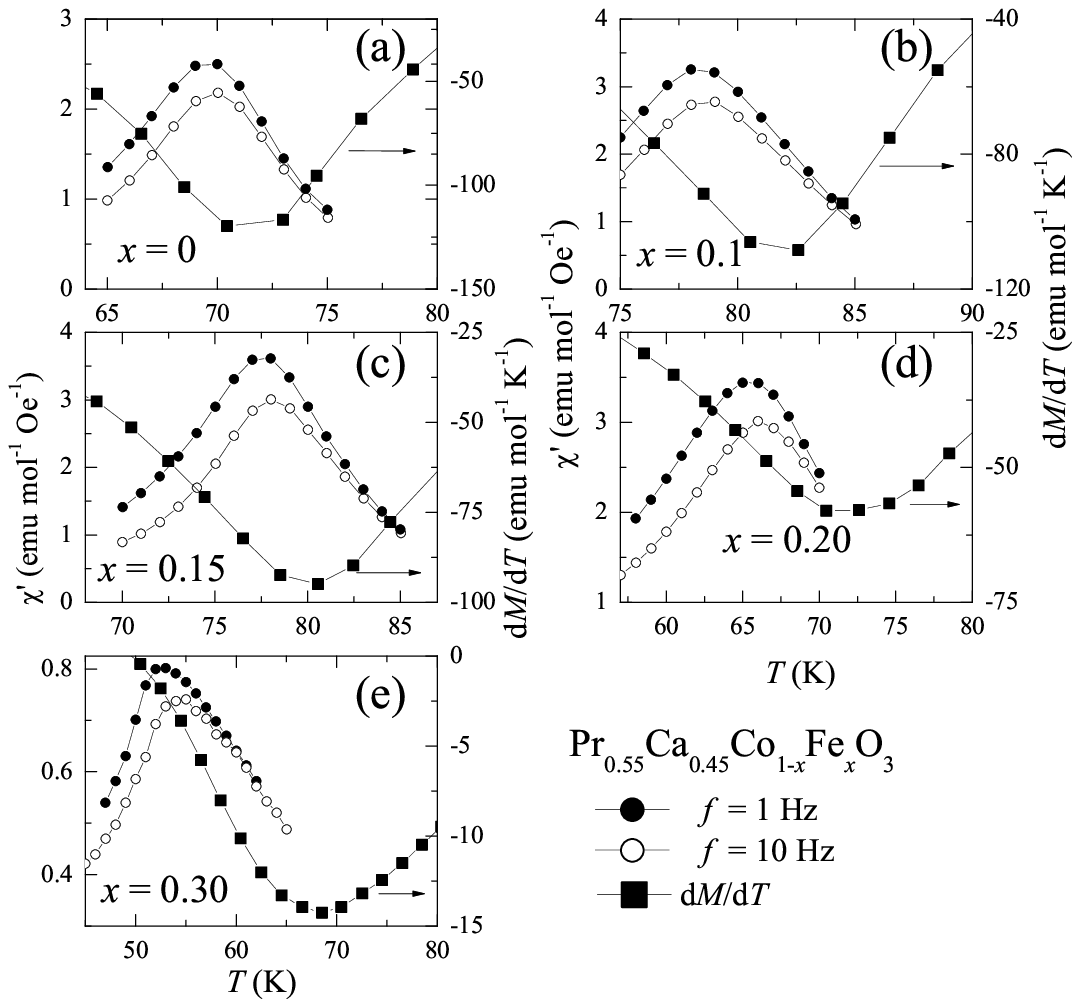}
\caption{Temperature dependence of in-phase ac susceptibility
$\chi'(T)$ at 1 Hz and 1 kHz measured in ac the magnetic field of
3.8 Oe for Pr$_{0.55}$Ca$_{0.45}$Co$_{1-x}$Fe$_{x}$O$_{3}$ with $x$
= 0, 0.1, 0.15, 0.2, and 0.3. As a comparison to dc data, the
d$M$/d$T$ is plotted together with the ac data. }
\end{figure}

\begin{figure}
\centering
\includegraphics[width=0.75\textwidth]{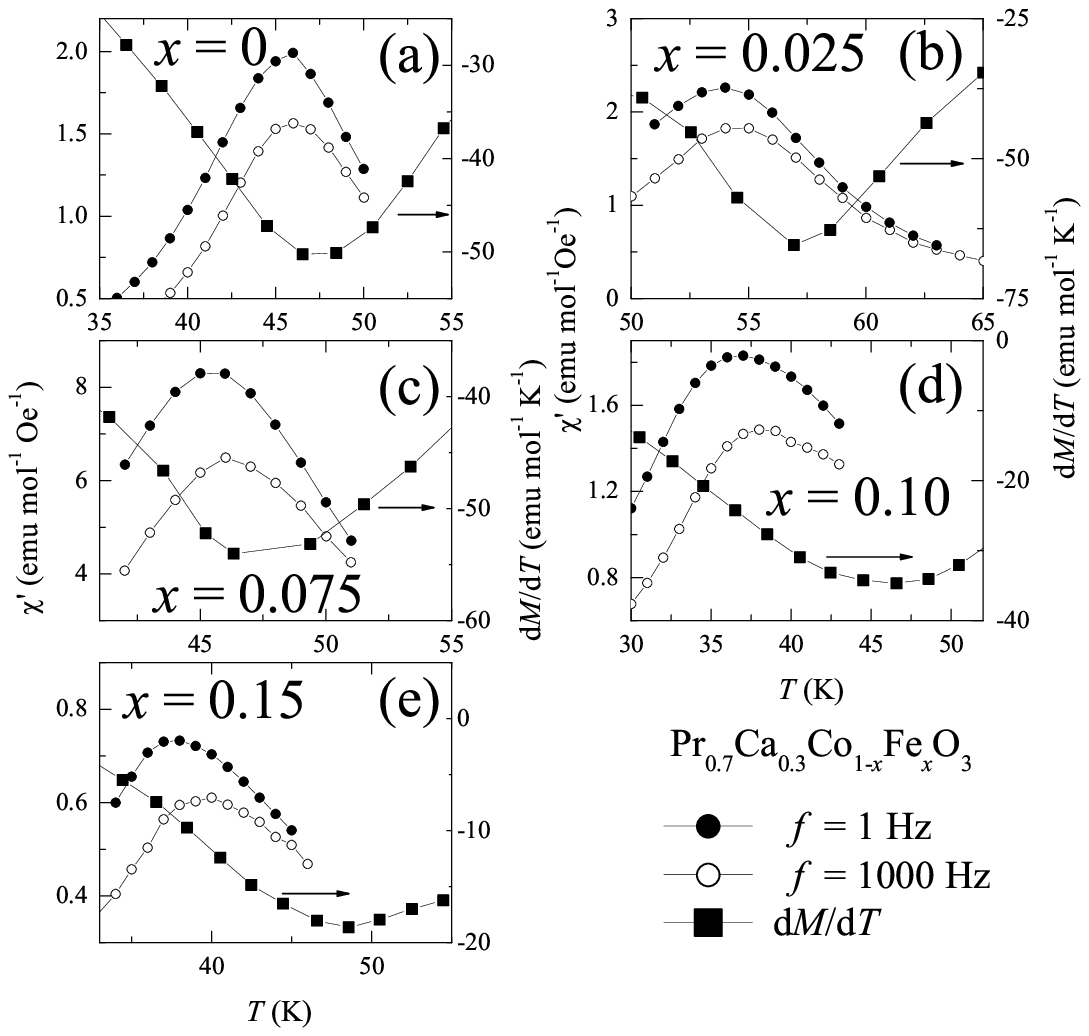}
\caption{Temperature dependence of in-phase ac susceptibility
$\chi'(T)$ at 1 Hz and 1 kHz measured in ac the magnetic field of
3.8 Oe for Pr$_{0.7}$Ca$_{0.3}$Co$_{1-x}$Fe$_{x}$O$_{3}$ with $x$ =
0, 0.025, 0.075, 0.1, and 0.15. As a comparison to dc data, the
d$M$/d$T$ is plotted together with the ac data.}
\end{figure}

\begin{figure} \centering
\includegraphics[width=0.8\textwidth]{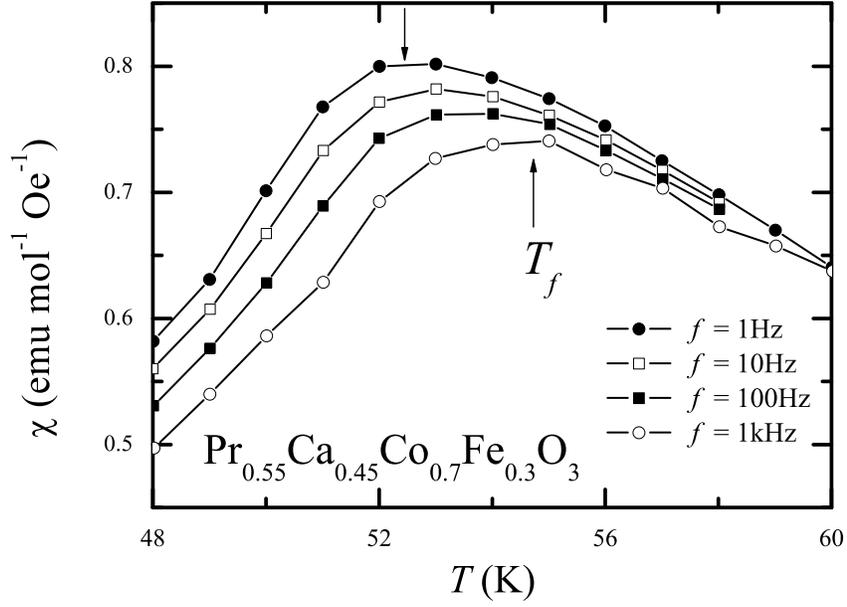}
\caption{Closeup of the temperature dependence of the in-phase ac
susceptibility for
Pr$_{0.55}$Ca$_{0.45}$Co$_{0.7}$Fe$_{0.3}$O$_{3}$, at four different
frequencies in the range 1 $-$ 1000 Hz.}
\end{figure}

\begin{figure}
\centering
\includegraphics[width=0.8\textwidth]{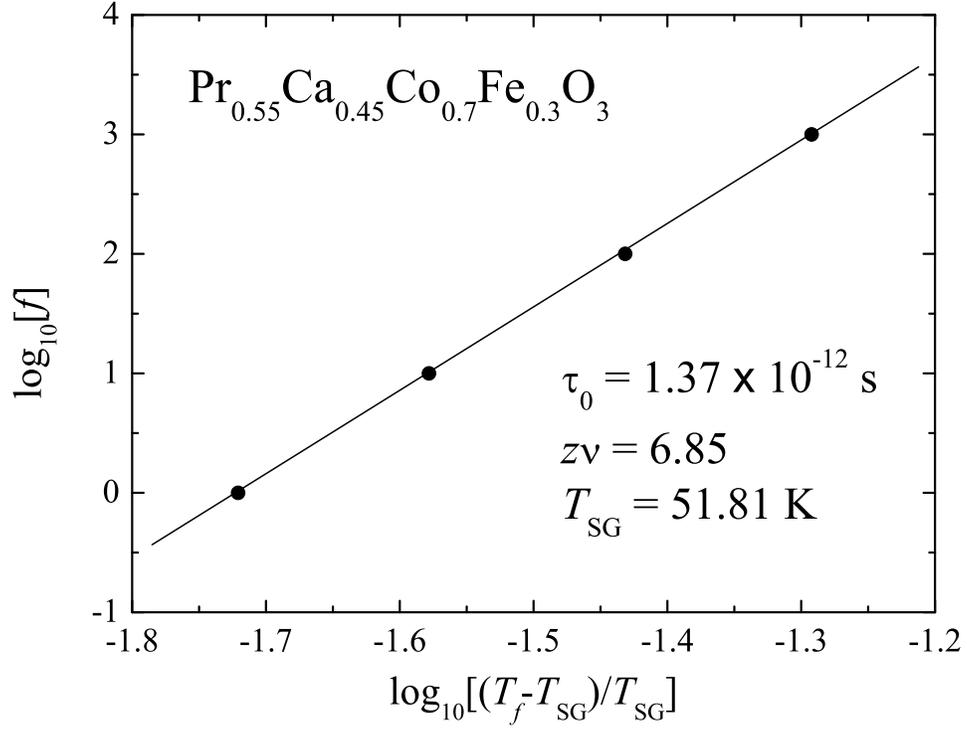}
\caption{log$_{10}$[$f$] vs log$_{10}$[($T_f$-$T_{SG}$)/$T_{SG}$]
for Pr$_{0.55}$Ca$_{0.45}$Co$_{0.7}$Fe$_{0.3}$O$_{3}$, demonstrating
good agreement with Eq. (1). $T_f$ is determined by the temperature
with d$\chi'(T)$/d$T$ = 0. The solid line is a best fit to the data
with the parameters shown in the figure.}
\end{figure}

\begin{figure}
\centering
\includegraphics[width=0.9\textwidth]{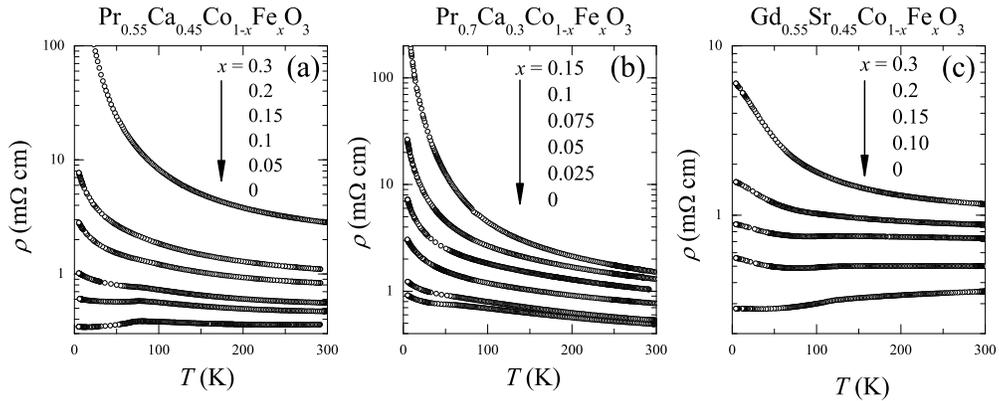}
\caption{Temperature dependence of resistivity for the
polycrystalline samples of: (a)
Pr$_{0.55}$Ca$_{0.45}$Co$_{1-x}$Fe$_{x}$O$_{3}$; (b)
Pr$_{0.7}$Ca$_{0.3}$Co$_{1-x}$Fe$_{x}$O$_{3}$; (c)
Gd$_{0.55}$Sr$_{0.45}$Co$_{1-x}$Fe$_{x}$O$_{3}$. }
\end{figure}

\begin{figure} \centering
\includegraphics[width=0.75\textwidth]{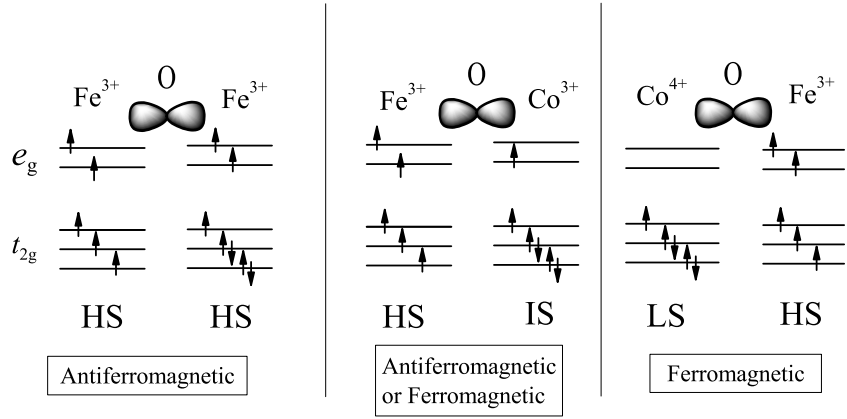}
\caption{Schematic model for the spin state configuration of
Fe$^{3+}$, Co$^{4+}$ and of Fe$^{3+}$, Co$^{3+}$ (left and right,
respectively). The oxygen orbitals are also drawn. HS, IS, and LS
represent high, intermediate and low spin, respectively.}
\end{figure}

\end{document}